\documentclass[%
 reprint,
 amsmath,amssymb,
 aps,
]{revtex4-2}

\usepackage{graphicx}% Include figure files
\usepackage{dcolumn}% Align table columns on decimal point
\usepackage{bm}% bold math
\usepackage{physics}
\usepackage{dsfont}
\usepackage{comment}
\usepackage{xcolor}

\begin{document}

\preprint{APS/123-QED}

\title{Flat Bands arising from Spin-Orbit Assisted Orbital Frustration}% Force line breaks with \\

\author{Zachariah Addison}%
\email{addison.64@osu.edu}
 \affiliation{Department of Physics, The Ohio State University, 191 West Woodruff Avenue, Columbus, Ohio 43210, USA}
\author{Nandini Trivedi}%
 \email{trivedi.15@osu.edu}
 \affiliation{Department of Physics, The Ohio State University, 191 West Woodruff Avenue, Columbus, Ohio 43210, USA}

\date{\today}

\begin{abstract}
We present general design principles for engineering and discovering periodic systems with flat bands.  Our paradigm exploits spin-orbit assisted orbital frustration on a lattice to produce band structures that contain multiplets of narrowly dispersing bands whose bandwidth is smaller than all other energy scales of the problem including the band gap surrounding the flat bands.  We present a series of models in 1D and 2D on various lattices with different intracellular spin-orbit like potentials that hybridize the degrees of freedom in the unit cell.  As an alternative to machine learning based exhaustive searches, these design principles and models can be used to search for flat band systems in a variety of physical settings and can be used to investigate the role of weakly dispersing highly orbitally frustrated degrees of freedom in systems where the interactions dominate over the kinetic energy scales of the system.
\end{abstract}

\maketitle

\section{Introduction}

Quantum materials continue to surprise us with fundamentally new emergent phases. Some of the most exciting recent developments include
 moiré materials with flat bands, often topological, where correlations are necessarily strong.
 
Usually the tunneling strength $t$ of electrons in materials leads to bands that naturally disperse with a bandwidth of order $W\sim t$.
Flat band systems are those for which the bandwidth is suppressed relative to the systems excitation gaps $\Delta$; such systems are rare.  Examples of flat bands can be found in the quantum Hall effect in a two-dimensional electron gas in a magnetic field, quantum anomalous Hall systems, periodic moiré structures by twisting layers in van der Waals coupled materials, and a few highly fine-tuned tight binding models on the lattice that are designed to lead to a subset of bands with exactly no dispersion, such as the Lieb lattice.  

Though the number of known completely dispersion-less systems are few (neglecting the trivial localized atomic insulators), the presence of flat bands has manifested in a multitude of exotic states of matter where interactions dominate and new topological phases emerge \cite{mielke1991ferromagnetic,mielke1993ferromagnetism, wu2007flat,huber2010bose,goda2006inverse,chalker2010anderson,ye2021flat}.  
Most theoretical constructions of flat-band systems have restricted the analysis to systems with a subset of completely flat bands in which the dispersion is completely absent \cite{lieb1989two,mielke1991ferromagnetism,mielke1992exact,mielke1992exact2,bergman2008band,green2010isolated,calder2010magnetic,liu2022orbital}.  However, in almost all materials only the ratio of the band width $W$ to all other energy scales of the system (including the systems band gaps) need be small in order to reveal the phenomena reliant on the existence of the flat-bands, thereby expanding the scope of possible materials.  A recent machine learning based study of flat band systems has used density function theory and the Inorganic Crystal Structure Database to catalogue over 2,000 such material candidates \cite{vergniory2019complete, vergniory2021all, li2021catalogue}.  As such, here we relax the condition of designing completely dispersion-less bands and present general design principles for achieving flat band systems with a subset of bands whose bandwidths are  much smaller than all other energy scales of the system.

Frustration has typically been used to indicate the inability to satisfy magnetic interactions locally because of competing interactions between local moments. Here we extend the concept of frustration to the band picture where the ability of electrons to tunnel on the lattice is hindered. We emphasize that this hindrance is not a consequence of an effectively large lattice constant as in the moire lattices, but arises due to {\em interference} between equivalent tunneling pathways. Specifically, the mechanism by which frustration arises here does not rely on large multi-site unit cells or on external fields, but instead relies solely on engineered orbital frustration that exploits multi-orbital spin-orbit assisted mixing on the lattice we dub {\it spin-orbit assisted orbital frustration}.

Spin-orbit assisted orbital frustration occurs when a large intra-cellular potential, $\lambda$, splits the degrees of freedom in a unit cell into multi-degenerate multiplets that are orthogonal to the kinetic processes that couple degrees of freedom between unit cells.  This orthogonality frustrates the kinetic processes $t$ on the lattice such that a degenerate multiplet is broken into a set of bands that disperse with bandwidth $W_F\sim t^2/\lambda$, that for large intra-cellular potential is small compared to bands in the absence of frustration that disperse canonically with bandwidths $W\sim t$.

\begin{figure*}[!htb]
    \centering
    \includegraphics[width=0.99\textwidth]{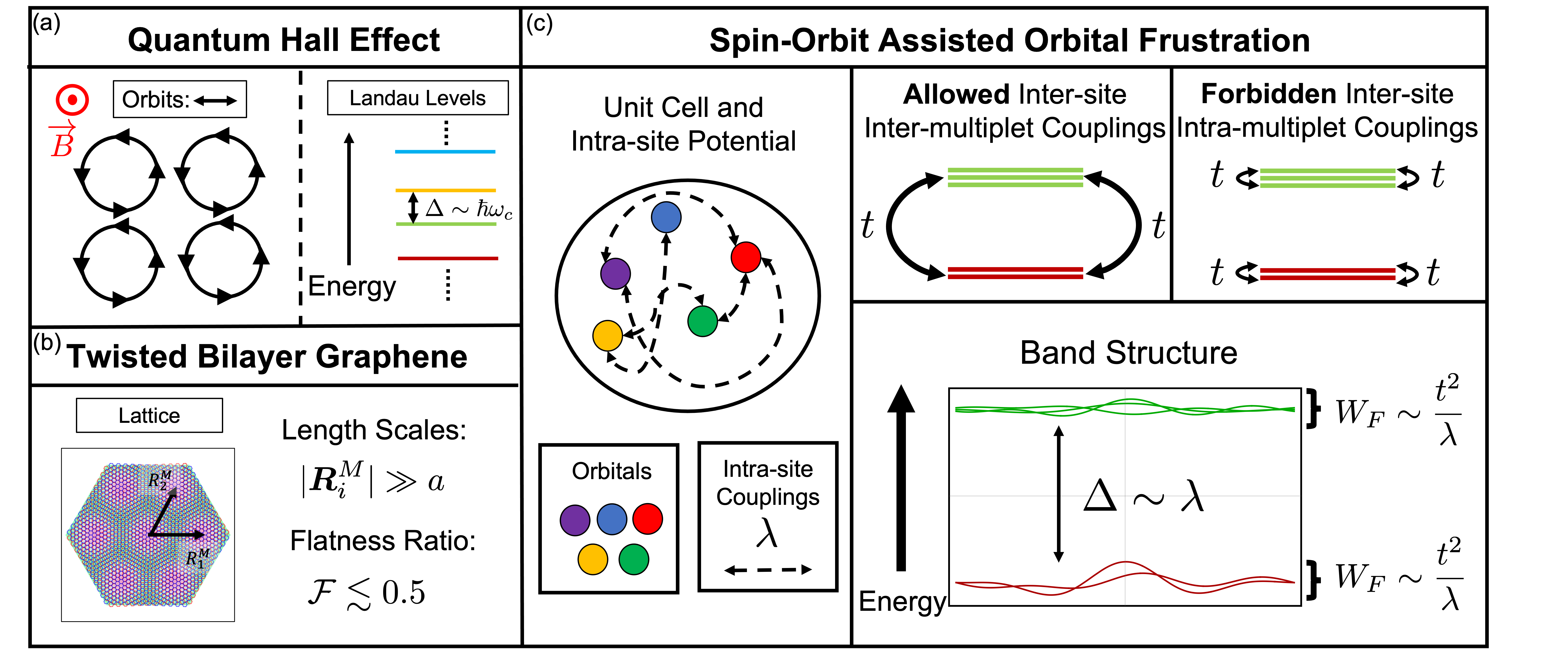}
    \caption{Various mechanisms leading to flat bands.  (a) In the quantum Hall effect a strong magnetic field forces the electron into quantized cyclotron orbits that leads to a Landau level spectrum of completely flat bands separated by an energy gap $\Delta=\hbar\omega_c$ proportional to the cyclotron frequency $\omega_c=eB/cm_e$. (b) For special angles of rotational misalignment in a stack of two graphene sheets a particular interlayer moiré potential develops of characteristic length $|\bm{R}^M_i|$ much larger than the intralayer graphene lattice constant $a$.  The potential quenches the kinetic energy of electrons in both layers and leads to flat bands in the Brillouin zone with flatness ratios $\mathcal{F}\lesssim 0.5$. (c) Spin-orbit assisted orbital frustrated systems exploit the intra-site potential, $\lambda$, that couples orbital degrees of freedom in each unit cell to engineer flat bands by allowing inter-site processes, $t$, that couple states in the different band multiplets of $\lambda$, while forbidding kinetic processes that couple states in the same band multiplets.  These systems exhibit orbital frustration that leads to frustrated flat band multiplets separated by energy gaps $\Delta\sim \lambda$ with bandwidths $W_F\sim t^2/\lambda$ and flatness ratios $\mathcal{F}\sim (t/\lambda)^2$.}
    \label{figSOAOF}
\end{figure*}

In systems where interactions dominate over kinetic couplings, the associated non-interacting theory at a particular energy has a large density of states and a large number of modes with small group velocities such that the kinetic energy is minimal.  Importantly the systems of interest should host multiplets of flat bands with bandwidths $W$ much smaller than the band gap $\Delta$ surrounding a particular multiplet leading to small flatness ratios $\mathcal{F}=W/\Delta$. Small $\mathcal{F}$ can lead to a fractionalization of a system's quanta in the presence of interactions, leading to emergent topological ordered phases that can show fractionalization of charge, fractionalization of spin, and quantized anomalous and topological Hall effects.  For example, in the fractional quantum hall effect a strong magnetic field forces the electrons into quantized circular orbits that describe dispersionless flat Landau levels (see Fig. \ref{figSOAOF}(a)) \cite{stormer1999fractional,jain1990theory,haldane1983fractional,wen1995topological}.  Likewise, strongly correlated superconducting and Mott insulating phases have recently been observed in twisted bilayer graphene where flat bands emerge from the presence of a large multi-orbital moiré potential whose characteristic size is much larger than the lattice constant of a single graphene sheet (see Fig. \ref{figSOAOF}(b))\cite{bistritzer2011moire,cao2018correlated,cao2018unconventional}.  The flat bands of spin-orbit assisted orbital frustration circumvents the need for large unit cells or strong external perturbing fields to generate flat bands by relying on deconstructive orbital mixing on the lattice to suppress bandwidths and lead to an enhanced flattness ratio $\mathcal{F}\sim (t/\lambda)^2$ (see Fig. \ref{figSOAOF}(c)).

Beyond the dispersion of bands in the Brilloiun zone another distinguishing and important characteristic of a band structure is the evolution of the Bloch eigenstates of the Hamiltonian with respect to their crystal momentum.  The Berry curvature and the quantum metric describe this geometric structure of the Bloch bundle, both of which lead to unique transport properties of the electron in the presence of external fields \cite{niu2010berry,holder2020consequences,mera2021khaler,xiao2019nonreciprocal,graf2021berry,gilles2016geometric}.  For example, in the presence of a homogenous electric field the electron acquires an anomalous velocity transverse to its momentum, while in the presence of heterogeneous electric fields a system's linear response couples directly to the quantum metric along the Fermi surface \cite{PhysRevB.99.121111}.  Spin-orbit assisted orbital frustration relies on the presence of inter-orbital intercell kinetic couplings that tends to mix the degrees of freedom in the unit cell.  In momentum space this twisting can result in exotic Berry curvature distributions in the Brilloiun zone when either time reversal or inversion symmetries are broken \cite{zhang2021orbital}.  Furthermore in systems that contain an orbitally frustrated multi-degenerate multiplet, small time reversal breaking perturbations that could arise in the presence of small interactions can lead to anomalous topological states endowed from the orbital mixing in the frustrated lattice.  The importance of these topological considerations in flat band systems have been studied in many perfectly dispersionless systems and in particularly for magic angle twisted bilayer graphene \cite{po2018origin,po2019faithful,song2019all,xie2020topology,chiu2020fragile,ma2020spin}.

Spin-orbit assisted orbital frustrated flat band systems can be engineered in arbitrary lattice structures in any dimension as long as the degrees of freedom in the unit cell is larger than one.  Below we detail examples of orbital frustration and emergent flat bands on the bipartite 1D chain, square, triangular, and honeycomb lattices.  These models provide a pathway to engineer flat bands in a variety of periodic systems with diverse intra-site and inter-orbital couplings.  Apart from electronic systems other highly tunable platforms, like cold atoms systems and photonic crystals, also provide a framework to study systems where this type of orbital frustration could be present resulting in flat bands and therfore dominant inter-multiplet interactions.

\section{Periodic Systems and Band Structures}

Periodic systems admit a discrete translation symmetry whereby the system is left invariant under translation by its set of lattice vectors.  In quantum mechanics the commutator of the Hamiltonian with the operators describing these translations vanish  such that the eigenstates of the Hamiltonian are Bloch modes: simultaneous eigenstates of these translation operators and indexed by a crystal momenta, $\bm{k}$, that describe the irreducible representations of the translation group of the system.  The band structure of a system describes the mapping from $\bm{k}$ to the energy eigenvalues of the Hamiltonian.  In classical mechanics any periodic system whose dynamics are describe by a linear differential operator can be written in terms of Bloch modes with momentum $\bm{k}$ taking values in the first Brillouin zone.  Fourier transform of the operator with respect to time leads to a generalized eigenvalue equation for the dispersion relations $\omega_n(\bm{k})$.  In both contexts the full eigenspectrum of the system is determined by the band energies and associated Bloch eigenvectors.

\begin{comment}
%%%%%%%%%%%%%%%
The properties of a system can be determined from knowledge of these Bloch eigenvalues and eigenstates, the latter of which is defined up to a momentum dependent phase.  The geometry and topology of the Bloch bundle describes the evolution of the phase of the Bloch eigenstates in the Brilloiun zone.  The nature of the Berry connection, curvature, and metric on this space is intimately tied to comprehending the ways in which the degrees of freedom in the state vector can be manipulated.  The topology of the Bloch bundle is encoded in integral forms for systems with isolated multiplets of bands whose bandgaps remain finite throughout the Brilloiun zone.  The region between two distinct topological involutions of the Bloch bundle can host localized energy eigenmodes.

On the other hand, the energy eigenvalues of a system determine the density of states and the group velocity of modes at a particular energy throughout the Brilloiun zone.  Symmetry can cause bands to touch at high symmetry momenta in the Brilloiun zone leading to topological structures of multifold degenerate eigenvalues.  Associated to these special band crossings points are topological numbers and surface Fermi arcs that describe topological modes confined to the system's surface.
%%%%%%%%%%%%%%%
\end{comment}

Here we describe our systems in a tight-binding framework whereby the degrees of freedom of the problem can be incorporated through local creation and annihilation operators $\hat{c}^\dagger_{i\alpha}$ and $\hat{c}_{i\alpha}$ that create and annihilation quanta in unit cell $i$ of orbital character $\alpha$ at position $\bm{R}_i+\bm{\tau}_\alpha$.  These operators span the state space or Hilbert space of the system and as such the single body non-interacting Hamiltonian can be written entirely in terms of these operators

\begin{equation}
    \hat{H}=\sum_{ij,\alpha\beta}t_{ij}^{\alpha\beta}\hat{c}^\dagger_{i\alpha}\hat{c}^{\phantom{\dagger}}_{j\beta}
\end{equation}

\noindent
For periodic systems $t^{\alpha\beta}_{ij}=t^{\alpha\beta}(\bm{R}_i-\bm{R}_j)$ and the eigenstates of the Hamiltonian are Bloch modes

\begin{equation}
    \ket{\Psi_n(\bm{k})}=\dfrac{1}{\sqrt{N}}\sum_{l\alpha}e^{i\bm{k}\cdot\hat{\bm{r}}}u_n^\alpha(\bm{k})\ket{\bm{R}_l,\bm{\tau}_\alpha}
\end{equation}

\noindent
that can be indexed by a band number $n$ and crystal momentum $\bm{k}$ taking values in the first Brillouin zone and that satisfy $\hat{H}\ket{\Psi_n(\bm{k})}=\varepsilon_n(\bm{k})\ket{\Psi_n(\bm{k})}$.  Here the state vectors $\ket{\bm{R}_i,\bm{\tau_\alpha}}$ describe the occupation of quanta of orbital character $\alpha$ at position $\bm{R}_i+\bm{\tau}_\alpha$: $\hat{c}^\dagger_{i\alpha}\ket{0}=\ket{\bm{R}_i,\bm{\tau}_\alpha}$, with $\ket{0}$ being the vacuum state.  The eigenstates and eigenvalues are determined by finding the periodic part of the Bloch eigenstates $u_n^\alpha(\bm{k})=\braket{\alpha}{u_n(\bm{k})}$ that satisfy

\begin{equation}
     \hat{H}(\bm{k})\ket{u_n(\bm{k})}=\varepsilon_n(\bm{k})\ket{u_n(\bm{k})}
     \label{evaleq}
\end{equation}

\noindent
where the Bloch Hamiltonian $\hat{H}(\bm{k})$ is determined from knowledge of the tight-binding coefficients $t^{\alpha\beta}(\bm{\delta})$

\begin{equation}
    \bra{\alpha}\hat{H}(\bm{k})\ket{\beta}=\sum_{\delta}e^{-i\bm{k}\cdot\bm{\delta}}e^{-i\bm{k}\cdot(\bm{\tau}_\alpha-\bm{\tau}_\beta)}t^{\alpha\beta}(\bm{\delta})
    \label{BlochHam}
\end{equation}

In general, systems that permit a band structure are described by a linear differential operator $\hat{\mathcal{O}}(\partial_{r_i},\partial_t)$ whose action on the system's state vector $\bm{v}(\bm{r},t)$ lead to a generalized eigenvalue equation for the Bloch modes

\begin{equation}
    \hat{\mathcal{O}}(\partial_{r_i}+ik_i,\omega_n)\bm{v}_n(\bm{r},\bm{k})=0
    \label{classicalH}
\end{equation}

\noindent
whose solutions $\omega_n\rightarrow \omega_n(\bm{k})$ determine the band energy eigenvalues of the system \cite{collet2011floquet,joannopoulos2011photonic,kushwaha1996classical}.  These models have been used to analyze flat bands in optical lattices and superconducting circuits \cite{baboux2016bosonic,kollar2019hyperbolic,leykam2018artificial}.  For example, the study of electromagnetic waves in linear dielectric materials leads to a generalized eigenvalue equation for the electric and magnetic field.  The harmonic transverse magnetic modes of a simple 2D linear dielectric material are governed by the equations

\begin{gather}
    \dfrac{1}{\varepsilon({\bm{r}})}\bm{\nabla}\times\bm{\nabla}\times \bm{E}(\bm{r},\omega)-\dfrac{\omega^2}{c^2}\bm{E}(\bm{r},\omega)=0 \nonumber \\
    \bm{H}(\bm{r},\omega)+\dfrac{i}{\mu_0\omega}\bm{\nabla}\times \bm{E}(\bm{r},\omega)=0
\end{gather}

\noindent
where connection with equation \ref{classicalH} is made by expanding the electromagnetic fields in Bloch modes with momenta $\bm{k}$.  In practice and for numeric calculation these equations are usually discretized and can be recast into a framework similiar to the tightbinding description used above \cite{wu2015scheme,wang2020topological}.  For these reasons and the natural applicability to describe quantum mechanical processes we will adopt the tightbinding framework for the rest of this manuscript.

In the absence of intercell tight-binding coefficients, $t^{\alpha\beta}(\bm{\delta})=0$ for $\bm{\delta}\neq \bm{0}$, the eigenvalues of the Bloch Hamiltonian are momentum independent: $\varepsilon_n(\bm{k})\rightarrow \varepsilon_n$.  This can be seen by making a momentum dependent unitary gauge transformation on the Bloch Hamiltonian.

\begin{equation}
    \hat{U}(\bm{k})\hat{H}(\bm{k})\hat{U}^\dagger(\bm{k})=\hat{H}'(\bm{k})
\end{equation}

\noindent
with $\bra{\alpha}\hat{U}(\bm{k})\ket{\beta}=e^{i\bm{k}\cdot\bm{\tau}_\alpha}\delta_{\alpha\beta}$.  Using equation \ref{BlochHam} we find

\begin{equation}
    \hat{H}'(\bm{k})=\sum_{\delta,\alpha\beta}e^{-i\bm{k}\cdot\bm{\delta}}t^{\alpha\beta}(\bm{\delta})\ket{\alpha}\bra{\beta}
\end{equation}

\noindent
which in the limit of purely intracell tight-binding coefficients is a momentum independent operator with momentum independent eigenvalues, $\varepsilon_n$.

Dispersion in the band structure then arises from the strength and character of the intercell hopping elements coupling degrees of freedom between the unit cells of a lattice.  To study the dispersion of bands we can decompose our original Hamiltonian into terms describing intracell $\hat{H}_\text{intra}$ and intercell hopping $\hat{H}_\text{inter}$:

\begin{gather}
    \hat{H}=\hat{H}_{\text{intra}}+\hat{H}_{\text{inter}} \nonumber \\ \nonumber\\
    \hat{H}_{\text{intra}}=\sum_{i,\alpha\beta} \Lambda^{\alpha\beta}_i \hat{c}^\dagger_{i\alpha}\hat{c}^{\phantom{\dagger}}_{i\beta} \nonumber\\ \hat{H}_{\text{inter}}=\sum_{ij,\alpha\beta} \Gamma^{\alpha\beta}_{ij} \hat{c}^\dagger_{i\alpha}\hat{c}^{\phantom{\dagger}}_{j\beta}
\end{gather}

\noindent
with $\Gamma^{\alpha\beta}_{ii}=0$.  Due to the translation symmetry of the crystal $\Lambda^{\alpha\beta}_i=\Lambda^{\alpha\beta}$ independent of the unit cell $\bm{R}_i$ and $\Gamma^{\alpha\beta}_{ij}=\Gamma^{\alpha\beta}(\bm{R}_i-\bm{R}_j)$.  Similarly for the Bloch Hamiltonian we have

\begin{equation}
    \hat{H}(\bm{k})=\hat{H}_{\text{intra}}(\bm{k})+\hat{H}_{\text{inter}}(\bm{k}) 
\end{equation}

\noindent
Here we are interested in the energy eigenvalues of $\hat{H}(\bm{k})$ and thus for simplicity we instead focus on diagonalizing

\begin{gather}
\hat{H}'(\bm{k})=\hat{H}'_{\text{intra}}+\hat{H}'_{\text{inter}}(\bm{k}) \nonumber \\ \nonumber\\
\hat{H}'_{\text{intra}}=\hat{U}(\bm{k})\hat{H}_{\text{intra}}(\bm{k})\hat{U}^\dagger(\bm{k}) \nonumber \\
\hat{H}'_{\text{inter}}(\bm{k})=\hat{U}(\bm{k})\hat{H}_{\text{inter}}(\bm{k})\hat{U}^\dagger(\bm{k})
\end{gather}

\noindent
In the orbital basis this takes the form

\begin{gather}
    \bra{\alpha}\hat{H}'_{\text{intra}}\ket{\beta}=\Lambda^{\alpha\beta} \nonumber \\
    \bra{\alpha}\hat{H}'_{\text{inter}}(\bm{k})\ket{\beta}=\sum_\delta e^{-i\bm{\delta}\cdot\bm{k}}\Gamma^{\alpha\beta}(\bm{\delta})
    \label{BlochHam2}
\end{gather}

\noindent
By investigating the relationship between eigenfunctions of $\hat{H}_{\text{intra}}(\bm{k})$ and $\hat{H}_{\text{inter}}(\bm{k})$ we can determine whether a system exhibits spin-orbit assisted orbital frustration and whether a system will posses narrowly dispersing frustrated band multiplets.

\section{intracell Multiplet Structures}
\label{multiplet}

We begin by analyzing the flat bands that can arise in the absence of intercell hopping.  In electronic systems the predominant intracell contribution to the Hamiltonian derive from the local crystal field potential and spin-orbit interactions that split a number of orbital degrees of freedom in each unit cell into a multiplet structure of degenerate manifolds of Bloch states. Upon the inclusion of intercell processes in $\hat{H}$ these degenerate states hybridize renormalizing the flat bands leading to dispersion across the Brilloiun zone.  For example in an octahedral environment atomic $d$-orbitals split into a threefold degenerate $t_{2g}$ multiplet of $d_{xy}$, $d_{yz}$ and $d_{xz}$ orbital states and a two fold degenerate $e_g$ multiplet of $d_{x^2-y^2}$ and $d_{z^2}$, whereas in a square planar environment atomic orbitals split into a two fold degenerate multiplet of $d_{xy}$ and $d_{yz}$ orbitals and three singly degenerate multiplets of $d_{z^2}$, $d_{xy}$, and $d_{x^2-y^2}$ orbitals.  With the inclusion of spin-orbit interaction these multiplets get further split.  For example for $d$-orbitals in the octahedral environment the three fold degenerate $t_{2g}$ multiplet gets split into a two fold degenerate effective $J=1/2$ manifold of states and an four fold degenerate effective $J=3/2$ manifold of states.

Each degenerate multiplet is spanned by the eigenstates of $\hat{H}_{\text{intra}}$ or $\hat{H}'_{\text{intra}}$.  The eigenstates of the latter are describe by the vectors

\begin{equation}
    \ket{u_n^{\text{intra}}}=\sum_\alpha \gamma^\alpha_n \ket{\alpha}
\end{equation}

\noindent
For example for the $d$ orbitals of the $t_{2g}$ multiplet the two degenerate states of the $J=1/2$ sector are spanned by the vectors

\begin{align}
    \ket{u_1^{J=1/2}}&= -\dfrac{1}{\sqrt{3}}\ket{d_{yz},\uparrow}+\dfrac{i}{\sqrt{3}}\ket{d_{zx},\uparrow}+\dfrac{1}{\sqrt{3}}\ket{d_{xy},\downarrow} \nonumber \\
    \ket{u_2^{J=1/2}}&=\dfrac{1}{\sqrt{3}} \ket{d_{yz},\downarrow}+\dfrac{i}{\sqrt{3}}\ket{d_{zx},\downarrow}+\dfrac{1}{\sqrt{3}}\ket{d_{xy},\uparrow}
    \label{eigstates1}
\end{align}

\noindent
While for the $d$ orbitals of the $t_{2g}$ multiplet the four degenerate states of the $j_{3/2}$ sector are spanned by the vectors

\begin{align}
    \ket{u_1^{J=3/2}}&=\dfrac{1}{\sqrt{2}}\ket{d_{yz},\uparrow}+\dfrac{1}{\sqrt{2}}\ket{d_{xy},\downarrow} \nonumber \\
    \ket{u_2^{J=3/2}}&=-\dfrac{1}{\sqrt{2}}\ket{d_{yz},\downarrow}+ \dfrac{1}{\sqrt{2}}\ket{d_{xy},\uparrow}\nonumber \\
    \ket{u_3^{J=3/2}}&=\dfrac{i}{\sqrt{6}}\ket{d_{yz},\downarrow}+\sqrt{\dfrac{2}{3}}\ket{d_{zx},\downarrow}+\dfrac{i}{\sqrt{6}}\ket{d_{xy},\uparrow} \nonumber \\
    \ket{u_4^{J=3/2}}&=-\dfrac{i}{\sqrt{6}}\ket{d_{yz},\uparrow}+\sqrt{\dfrac{2}{3}}\ket{d_{zx},\uparrow}+\dfrac{i}{\sqrt{6}}\ket{d_{xy},\downarrow}
    \label{eigstates2}
\end{align}

\noindent
These states can be determined by diagonalizing the effective onsite spin-orbit interaction for the $t_{2g}$ manifold

\begin{equation}
    \Lambda^{\alpha\beta}=\lambda_{so} \bra{\alpha} \hat{\bm{L}}\cdot \hat{\bm{S}} \ket{\beta}
    \label{LScouple}
\end{equation}

\noindent
where $\lambda_{so}$ is the energy scale for intracellular spin-orbit interactions such that $\hat{\bm{S}}=\hat{\bm{\sigma}}$ and such that the eigenvalues of $\Lambda$ are $-2\lambda_{so}$ with multiplicity two and $\lambda_{so}$ with multiplicity four.

In general, in the absence of intercell hopping, the intracell potentials described by $\hat{H}_{\text{intra}}$ lead to a multiplet structure of $M$ sets of $N$-fold degenerate bands whose energy eigenvalues and eigenvectors we denote as $\varepsilon_{m,i}$ and $\ket{u_i^m}$ with $m=1,...,M$ and $i=1,..,N$, separated by an energy, $\Delta$, the order of the intracell hopping potentials $\lambda$.  Each set in $M$ spans a $N$-dimensional subspace of the Hilbert space such that the total number of bands $M \times N$ equals the number of degrees of freedom in the systems unit cell.

\begin{figure*}[!htb]
    \centering
    \includegraphics[width=0.99\textwidth]{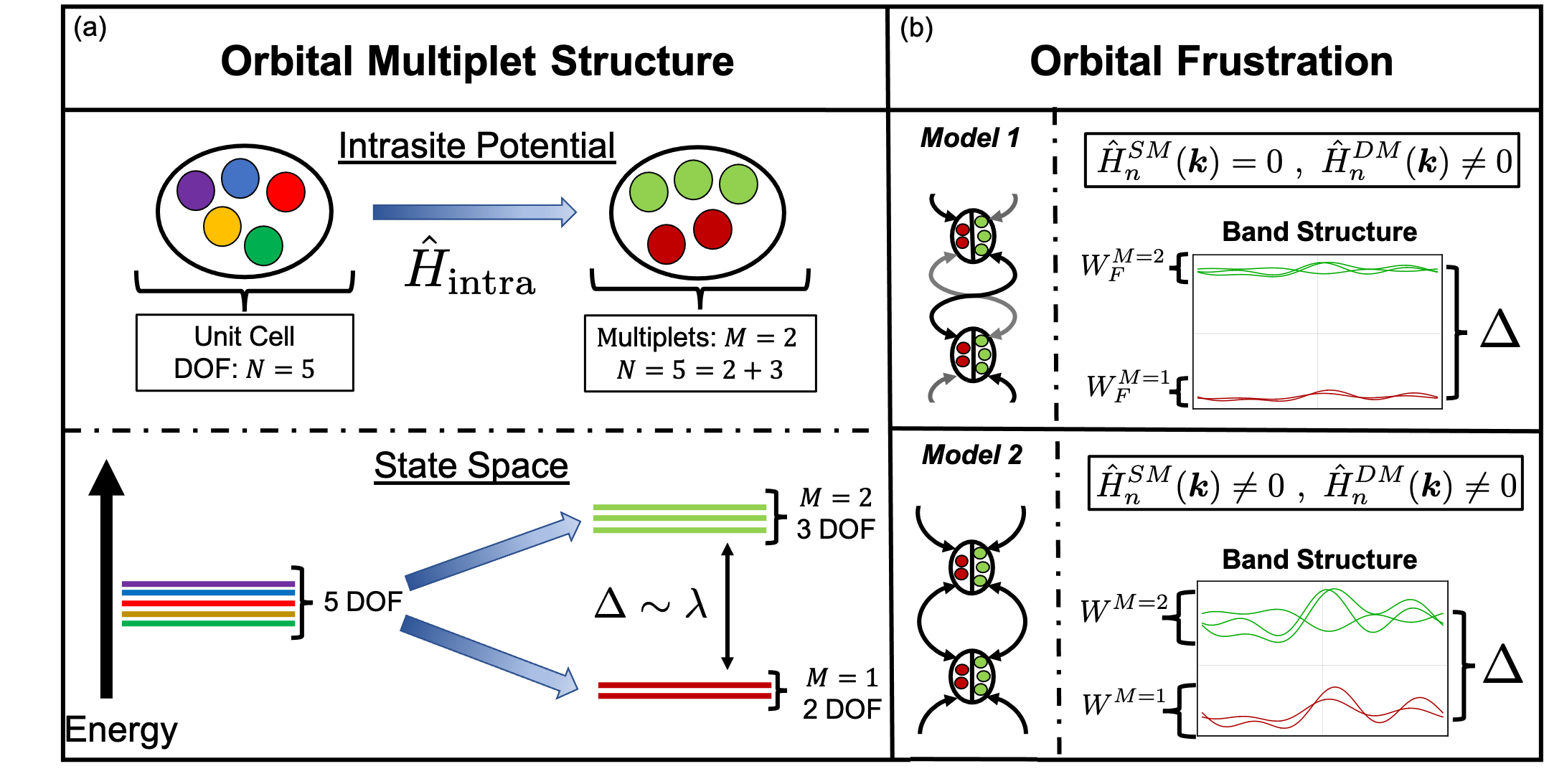}
    \caption{(a) Spin-orbit assisted orbital frustration relies on the presence of a large intra-site potential, $\hat{H}_{\text{intra}}\sim\lambda$, that couples the orbital degrees of freedom in the unit cell to produce a multiplet structure of degerate bands separated by an energy gap $\Delta\sim \lambda$. Here we show a unit cell of 5 degrees of freedom (DOF), split by an intra-site potential into two band multiplets: a multiplet ($M=1$) containing three degenerate eigenstates (green) of $\hat{H}_{\text{intra}}$ and a multiplet ($M=2$) containing two degenerate eigenstates (maroon) of $\hat{H}_{\text{intra}}$. (b) Orbital frustration requires the absence of inter-site intra-multiplet kinetic processes $\hat{H}_n^{SM}(\bm{k})=0$.  In {\it Model 1} inter-site intra-multipelt processes are forbidden, and only kinetic couplings between sites that couples states in different band multiplets are allowed leading to orbital frustrated flat bands with bandwidths $W_F\sim t^2/\lambda$.  In {\it Model 2} inter-site processes that couple states in the same band multiplet are allowed spoiling orbital frustration and leading to bandwidths $W\sim t$. }
    \label{fig1}
\end{figure*}

\section{Orbital Frustration}

Orbital frustration is the suppression of the bandwidths of multiplets of bands in a system's band structure deriving from the absence or limitation of specific hopping matrix elements in $\hat{H}_{\text{inter}}$.  In a general Hamiltonian and in the absence of orbital frustration the bandwidth of any given band in a band structure can be expected to be order the intercell hopping strength $t$. In the presence of strong intracell potentials $\lambda\gg t$ the set of bands splits into the multplet structures described in section \ref{multiplet} and for general intercell hopping the bandwidths of these multiplets will too be of order the intercell hopping strengths $t$.  This is most easily understood in the context of perturbation theory.

The ratio of $\lambda$ to $t$ is a small parameter by which a perturbative expansion of the energy eigenvalues in $l=t/\lambda$ of $\hat{H}(\bm{k})$ can be computed.  In a system with $M$ different multiplets, at zeroth order in $l$, the eigenvalues of $\hat{H}(\bm{k})$ are given by the momentum independent eigenvalues of $\hat{H}_{\text{intra}}$ which we write as $\xi_m$ with $m=1,...,M$.  In the basis of eigenstates of $\hat{H}'_{\text{intra}}$, $\ket{u_n^{\text{intra}}}$, the intracell Hamiltonian takes a block diagonal form of $M$, $N\times N$ diagonal matrices.

The first order correction to the energy eigenvalues $\xi_m$, $\varepsilon_{m,i}^{(1)}(\bm{k})$, is found by orthogonalizing the eigenstates of a given multiplet $\ket{u^m_i}$ with respect to $\hat{H}'_{\text{inter}}(\bm{k})$.  The first order corrections is determined by the eigenvalues of the matrix

\begin{equation}
    \mathcal{W}^m_{ij}(\bm{k})=\bra{u^m_i}\hat{H}'_{\text{inter}}(\bm{k})\ket{u^m_j}
\end{equation}

\noindent
The eigenvalues are of order the intercell hopping potentials $\varepsilon_{m,i}^{(1)}(\bm{k})\sim t$ and in general will lead to a hybridization of bands in a given multiplet breaking their degeneracy and leading to bandwidths $W\sim t$.

In frustrated systems the first order correction to the multiplet structure of $\hat{H}_{\text{intra}}$ vanishes, $\mathcal{W}^m_{ij}(\bm{k})=0$, for some $m\in M$.  These multiplets exhibit a reduction in bandwidth whose size is given by the next leading order correction to the eigenvalues of $\hat{H}_{\text{intra}}$

\begin{equation}
    \varepsilon^{(2)}_{m,i}(\bm{k})=\sum_{n\not\in m,j}\dfrac{\bra{u^m_i}\hat{H}'(\bm{k})\ket{u^n_j}\bra{u^n_j}\hat{H}'(\bm{k})\ket{u^m_i}}{\xi_m-\xi_n}
    \label{corr2}
\end{equation}

\noindent
These bands disperse with $\varepsilon^{(2)}_{m,i}(\bm{k})\sim l t$ leading to a suppression of the bandwidth $W/t$ of order $l$.

Given a set of onsite potentials $\hat{H}_{\text{intra}}$, orbital frustration is found by separating the allowed intercell kinetic hopping elements $\hat{H}'_{\text{inter}}(\bm{k})$ into inter-multiplet and intra-multiplet contributions.  The intercell inter-multiplet contributions $\hat{H}^{DM}_{nm}(\bm{k})$ describes the kinetic hopping between different multiplets, while the intercell intra-multiplet contribution $\hat{H}^{SM}_n(\bm{k})$ describes the kinetic hopping between the same multiplet.

\begin{equation}
    \hat{H}'_{\text{inter}}(\bm{k})=\sum_{n}\hat{H}^{SM}_n(\bm{k})+\sum_{n\neq m}\hat{H}^{DM}_{nm}(\bm{k})
\end{equation}

\noindent
First we define the projection operator onto a single multiplet

\begin{equation}
    \hat{\mathcal{P}}_n=\sum_i\ket{u_i^n}\bra{u^n_i}
\end{equation}

\noindent
where $i=1,...,N$, where $N$ is the number of state in the multiplet $n$.  The intercell intra-multiplet and inter-multiplet contribution can then be written as

\begin{align}
    \hat{H}_n^{SM}(\bm{k})&= \hat{\mathcal{P}}_n\hat{H}'_{\text{inter}}(\bm{k})\hat{\mathcal{P}}_n \nonumber\\
    \hat{H}^{DM}_{nm}(\bm{k})&= \hat{\mathcal{P}}_n\hat{H}'_{\text{inter}}(\bm{k})\hat{\mathcal{P}}_m \nonumber\\
\end{align}

\noindent
For systems in which $\hat{H}^{SM}_n(\bm{k})$ vanishes, $\mathcal{W}^n_{ij}(\bm{k})=0$ and orbital frustration forces the $n$-th multiplet to disperse with bandwidth $W\sim t^2/\lambda$.

It is useful to note that a momentum dependent projection of elements of a Bloch Hamiltonian, in principle, may lead 
%to a momentum dependent operator corresponding 
to long-ranged hopping processes.  However, in the construction proposed here the projection operators are {\it momentum independent} such that the lattice harmonic structure contained in $\hat{H}_n^{SM}(\bm{k})$ and $\hat{H}^{DM}_{nm}(\bm{k})$ are unchanged.  This is because the geometric character of the degrees of freedom in the unit cell remain independent of the lattice harmonic functions appearing in $\hat{H}_{\text{inter}}'(\bm{k})$. Hence the constraint $\mathcal{W}^n_{ij}(\bm{k})=0$ generically can be engineered in the absence of detailed knowledge of dependence of $\hat{H}_{\text{inter}}'(\bm{k})$ on the Bloch momenta.

\section{Models}
\label{models}

Here we present some simple toy models that demonstrate spin-orbit assisted orbital frustration.  We focus on unit cells with two, four, and six degrees of freedom coupled by intracellular potentials that lead to the multiplet structures described in section \ref{multiplet}.  We then determine the allowed and forbidden terms in $\hat{H}_{\text{inter}}$ that lead to the presence or absence of spin-orbit assisted orbital frustration.  

\subsection{Two Degrees of Freedom Per Site}

Consider a system with two degrees of freedom per site.  The general onsite Hamiltonian can be written as

\begin{equation}
    \hat{H}'_{\text{intra}}=\lambda_0\mathds{1}+\bm{\lambda}\cdot \hat{\bm{\sigma}}
\end{equation}

\noindent
where $\hat{\bm{\sigma}}=(\hat{\sigma}_x,\hat{\sigma}_y,\hat{\sigma}_z)$ are the two dimensional Pauli matrices.  The eigenvalues are $\xi_{\pm}=\pm|\bm{\lambda}|$ and the eigenvectors can be written as

\begin{align}
    \ket{u^-}&=(\sin(\theta/2)e^{-i\phi},-\cos(\theta/2)) \nonumber \\
    \ket{u^+}&=(\cos(\theta/2)e^{-i\phi},\sin(\theta/2))
\end{align}

\noindent
where $\bm{\lambda}=|\bm{\lambda}|(\sin(\theta)\cos(\phi),\sin(\theta)\sin(\phi),\cos(\theta))$.  The most general intercell hopping Hamiltonian can be written as 

\begin{equation}
    \hat{H}'_{\text{inter}}(\bm{k})=t_0(\bm{k})\mathds{1}+\bm{t}(\bm{k})\cdot \hat{\bm{\sigma}}
\end{equation}

\noindent
We can now construct $\mathcal{W}^{\pm}(\bm{k})$ in terms of $\bm{\lambda}$ and $\bm{d}(\bm{k})$.

\begin{align}
    \mathcal{W}^\pm(\bm{k})&=\pm t_z(\bm{k})\cos(\theta) \pm \sin(\theta)\bigg(t_x(\bm{k})\cos(\phi).  \nonumber \\
    &+t_y(\bm{k})\sin(\phi)\bigg)+t_0(\bm{k}) \nonumber \\
    &=\pm\dfrac{\bm{\lambda}}{|\bm{\lambda}|}\cdot \bm{t}(\bm{k})+t_0(\bm{k})
    \label{w2dof}
\end{align}

\noindent
Orbital frustration and band flattening will occur when either $\mathcal{W}^+(\bm{k})$ or $\mathcal{W}^-(\bm{k})$ vanishes.  

Take the simple example of $\bm{\lambda}=(0,0,\lambda_z)$ and $t_0=0$.  Then $\mathcal{W}^{\pm}(\bm{k})=\pm t_z(\bm{k})$ and orbital frustration occurs in the absence of intercell hopping terms in the Hamiltonian proportional to $\hat{\sigma}_z$.  This is simply understood by looking at the eigenstates of the intracell Hamiltonian for this specific onsite potential.  The eigenstates of $\hat{H}_{\text{intra}}$ are eigenstate of $\hat{\sigma}_z$, $\ket{u^+}=(1,0)$ and $\ket{u^-}=(0,1)$.  In order to induce hopping between the manifolds spanned by the $\ket{u^\pm}$ states, an intercell hop must flip the spinor index such that $\hat{H}'_{\text{inter}}(\bm{k}) \ket{u^\pm}\sim\ket{u^\mp}$.  The operators that allow for such a spinor flip are proportional to $\hat{\sigma}_x$ and $\hat{\sigma}_y$.  Thus in the absence of $\hat{\sigma}_z$, orbital frustration will develop. 

For arbitrary onsite potential the presence  of orbital frustration occurs when $\bm{t}(\bm{k})$ is orthogonal to $\bm{\lambda}$ and $t_0(\bm{k})=0$.  This can be understood in a similiar manner as the above simple example by performing a rotation of the coordinate system such that $\bm{\tilde{\lambda}}=\mathcal{R}\bm{\lambda}=(0,0,\tilde{\lambda}_z)$.  This is achieved by making a unitary transformation on the Hamiltonian such that 

\begin{align}
    \hat{U}(\bm{n}_\Omega,\Omega)\hat{H}'_{\text{inter}}\hat{U}^\dagger(\bm{n}_\Omega,\Omega) &=\hat{U}(\bm{n}_\Omega,\Omega)\bm{\lambda}\cdot\hat{\bm{\sigma}}\hat{U}^\dagger(\bm{n}_\Omega,\Omega) \nonumber \\
    &=\tilde{\lambda}_z\hat{\sigma}_z
    \label{rotation}
\end{align}

\noindent
where 

\begin{equation}
    \hat{U}(\bm{n}_\Omega,\Omega)=e^{-i\Omega\bm{n}_\Omega\cdot\hat{\bm{\sigma}}/2}=\cos(\Omega/2)-i\bm{n}_\Omega\cdot\hat{\bm{\sigma}}\sin(\Omega/2)
\end{equation}

\noindent
is the unitary transformation parameterized in terms of an axis of rotation $\bm{n}_\Omega=\bm{\lambda}/|\/\bm{\lambda}|\times (0,0,1)$ of unit norm and an angle $\Omega=\arccos(\lambda_z/|\bm{\lambda}|)$ such that equation \ref{rotation} is satisfied.  One can then determine the necessary operators for orbital frustration by making the corresponding unitary transformation of the Pauli operators that determine the presence of or absence of orbital frustration.  It follows that the allowed and forbidden terms in $\hat{H}'_{\text{inter}}(\bm{k})$ to achieve orbital frustration are

\begin{gather}
\text{Allowed}=\{\hat{U}^\dagger(\bm{n}_\Omega,\Omega)\hat{\sigma}_x\hat{U}(\bm{n}_\Omega,\Omega),\hat{U}^\dagger(\bm{n}_\Omega,\Omega)\hat{\sigma}_y\hat{U}(\bm{n}_\Omega,\Omega)\} \nonumber \\
\text{Forbidden}=\{\hat{U}^\dagger(\bm{n}_\Omega,\Omega)\hat{\sigma}_z\hat{U}(\bm{n}_\Omega,\Omega)\}
\end{gather}

\begin{figure*}[!htb]
    \centering
    \includegraphics[width=0.99\textwidth]{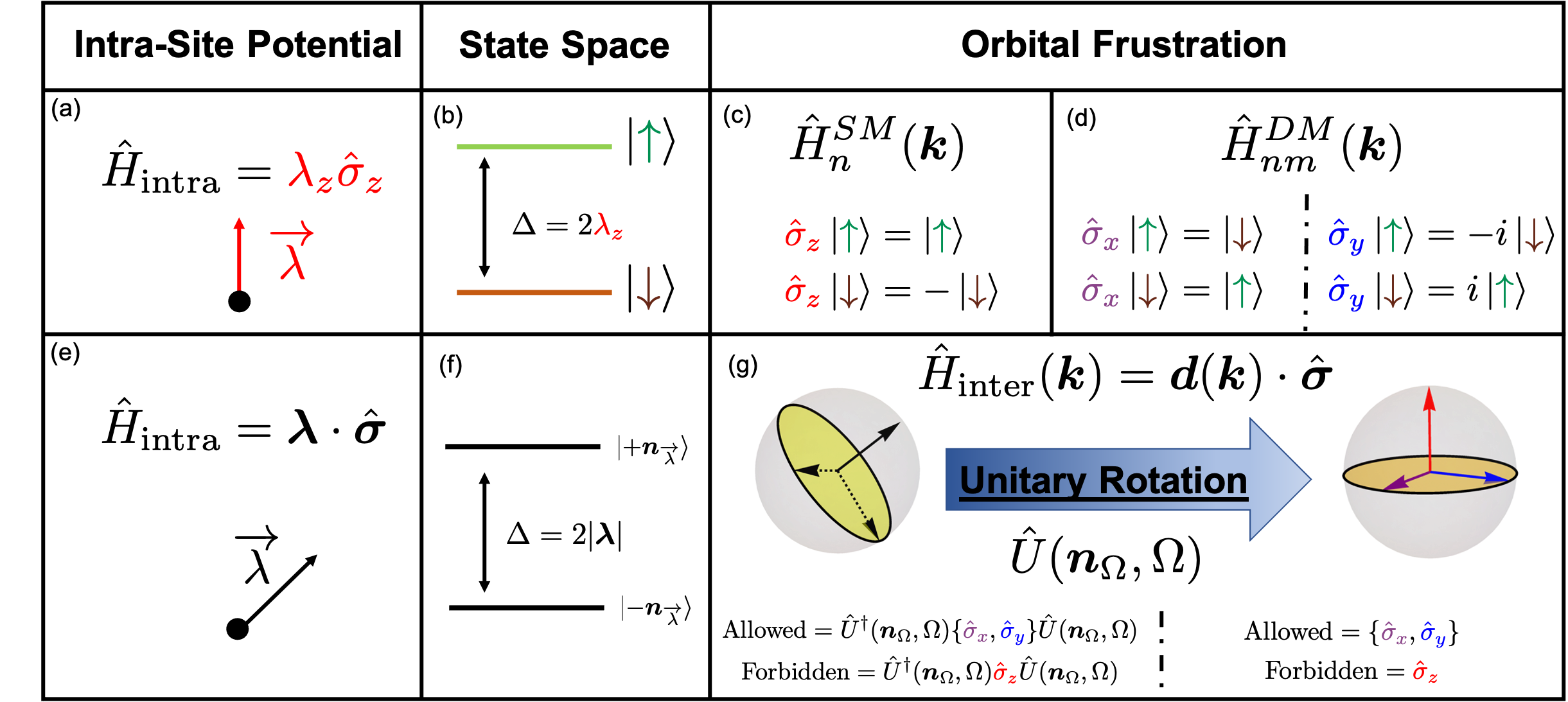}
    \caption{(a) Simple example of intra-site potential for a system with two degrees of freedom, ($\ket{\uparrow},\ket{\downarrow}$), per unit cell parameterized by orbital interaction $\lambda_z$. (b) Intra-site multiplet structure for intra-site potential (a).  States in the upper manifold (green) are eigenstates of $\hat{\sigma}_z$ with positive eigenvalue and states in the lower manifold (brown) are eigenstates of $\hat{\sigma}_z$ with negative eigenvalue. Inter-site processes can couple states within a given multiplet (c) or between multiplets (d).  In the absence of the former frustration will be present and lead to flat bands.  (e) General intra-site potential parameterized by $\bm{\lambda}$.  (f) Intra-site multiplet structure consist of two band multiplets.  States in the upper and lower multiplet are eigenstates of $\hat{H}_{\text{intra}}$ with eigenvalues $\pm |\bm{\lambda}|$. (g) Unitary rotation, $U(\bm{n}_\Omega,\Omega)$, of the Hamiltonian can transform the general intra-stie potential in (e) to the simpler potential described in (a).  Knowledge of $U(\bm{n}_\Omega,\Omega)$ determines the allowed and forbidden terms in $\hat{H}_{\text{inter}}(\bm{k})$ that can lead to the presence or absence of orbital frustration.}
    \label{fig2DOF}
\end{figure*}

\noindent
We see that in this model for general $\bm{\lambda}$ and $t_0(\bm{k})=0$, orbital frustration occurs in the presence of terms in the intercell Hamiltonian of the form $(\bm{t}(\bm{k})\times \bm{\lambda})\cdot\hat{\bm{\sigma}}$ for arbitrary $\bm{t}(\bm{k})$ and in the absence of terms in the intercell Hamiltonian proportional to $\bm{\lambda}\cdot\hat{\bm{\sigma}}$ (see Fig. \ref{fig2DOF}).

\subsection{Two $s=1/2$ Degrees of Freedom}
\label{4DOF}

Here we consider a model with two effective $s=1/2$ degrees of freedom per unit cell interacting with an onsite potential of the form

\begin{equation}
    \bra{\alpha}\hat{H}'_{\text{intra}}\ket{\beta}=\lambda_0\sum^3_{i=1}(\sigma_i\otimes\sigma_i)_{\alpha\beta}
    \label{intra4dof}
\end{equation}

\noindent
where $\otimes$ denotes the Kronecker product and $\alpha,\beta=1,...,4$.  The eigenstates of this onsite potential can be indexed by total angular momentum quantum numbers $\ket{J,m_J}$.  They split into two multiplets of states.  The triplet states with $J=1$

\begin{align}
    \ket{1,1}&=\ket{\uparrow\uparrow} \nonumber \\
    \ket{1,0}&=\dfrac{1}{\sqrt{2}}(\ket{\uparrow\downarrow}+\ket{\downarrow\uparrow}) \nonumber \\
    \ket{1,-1}&=\ket{\downarrow\downarrow}
\end{align}

\noindent
and eigenvalues $\xi_{J=1}=-3\lambda_0$, and the singlet state with $J=0$

\begin{equation}
    \ket{0,0}=\dfrac{1}{\sqrt{2}}(\ket{\uparrow\downarrow}-\ket{\downarrow\uparrow})
\end{equation}

\noindent
and eigenvalue $\xi_{J=0}=\lambda_0$.  We can now decompose the intercell Hamiltonian into terms described by the Kronecker product of two Pauli matrices

\begin{equation}
    \bra{\alpha}\hat{H}'_{\text{intra}}\ket{\beta}=\sum_{i=0}^3\sum_{j=0}^3 t_{ij}(\bm{k})(\sigma_i\otimes\sigma_j)_{\alpha\beta}
    \label{decomp4}
\end{equation}

\noindent
where $\sigma_0=\mathds{1}$.  The existence or absence of orbital frustration can be determined by constructing the matrices $\mathcal{W}^J_{ij}(\bm{k})$.  For $J=0$ we find

\begin{equation}
    \mathcal{W}^{J=0}(\bm{k})=\sum_{i=0}^3t_{ii}(\bm{k})
    \label{w4dof}
\end{equation}

\noindent
such that the allowed/forbidden terms in the intercell Hamiltonian that will lead to the presence/absence of orbital frustration are

\begin{align}
  \text{Allowed}=&\{\sigma_0\otimes\sigma_1,\sigma_0\otimes\sigma_2,\sigma_0\otimes\sigma_3, \nonumber \\
    &\sigma_x\otimes\sigma_0,\sigma_x\otimes\sigma_y,\sigma_x\otimes\sigma_z, \nonumber \\
    &\sigma_y\otimes\sigma_0,\sigma_y\otimes\sigma_x,\sigma_y\otimes\sigma_z \nonumber \\
    &\sigma_z\otimes\sigma_0,\sigma_z\otimes\sigma_x,\sigma_z\otimes\sigma_y\} \nonumber \\
    \text{Forbidden}=&\{\sigma_0\otimes\sigma_0,\sigma_x\otimes\sigma_x,\sigma_y\otimes\sigma_y,\sigma_z\otimes\sigma_z\}
    \label{halfhalfOF}
\end{align}

\noindent
In the absence of fine tuned linear combinations of $t_{ij}(\bm{k})$, the matrix $\mathcal{W}_{ij}^{J=1}$ for the triplet multiplet vanishes only when $t_{ij}(\bm{k})=0$.  However, for terms in the intercell Hamiltonian proportional to any of the allowed operators in equation \ref{halfhalfOF}, $\mathcal{W}^{J=1}$ contains a single vanishing eigenvalue.

\begin{align}
    \mathcal{W}_{nm}^{J=1}(\bm{k})&=\sum_{\alpha\beta}\braket{u^{J=1}_n}{\alpha}\sum_{i\neq j}t_{ij}(\bm{k})(\sigma_i\otimes \sigma_j)_{\alpha\beta}\braket{\beta}{u^{J=1}_m} \nonumber \\
   & \implies \text{Det}(\mathcal{W}^{J=1}(\bm{k}))=0
\end{align}

\noindent
For this choice of intercell Hamiltonian the three degenerate triplet states will hybridize such that the bandwidths of two of the three bands will be of order $t$, while the bandwidth of one of the three bands will be of order $t^2/\lambda$.  While one band will exhibit orbital frustration, in general the other two bands will be in the neighborhood of the flat band such that its flatness ratio is large.

In order to engineer terms in $\hat{H}'_{\text{inter}}(\bm{k})$ such that orbital frustration in both multiplets would be preserved, we begin by writing $\hat{H}^{DM}(\bm{k})$ in terms of the eigenstates of the $J=1$ and $J=0$ multiplet and arbitrary functions $t_i(\bm{k})$.

\begin{equation}
    \hat{H}^{DM}(\bm{k})=\sum_{i=1}^3t_{i}(\bm{k}) \ket{u^{J=1}_i}\bra{u^{J=0}}+h.c.
\end{equation}

\noindent
To make connection with the decomposition in equation \ref{decomp4} we separate $t_{i}(\bm{k})$ into its real $\bar{t}_i(\bm{k})$ and imaginary $\tilde{t}_i(\bm{k})$ parts such that

\begin{equation}
    \hat{H}^{DM}(\bm{k})=\sum_{i=1}^3\bigg(\bar{t}_{i}(\bm{k})\hat{S}_i+\tilde{t}_{i}(\bm{k})\Hat{D}_i\bigg)
\end{equation}

\noindent
where $\hat{S}_i$ and $\hat{d}_i$ can be written as

\begin{align}
    \hat{S}_i&=\ket{u^{J=0}_i}\bra{u^{J=1}}+\ket{u^{J=1}}\bra{u^{J=0}_i} \nonumber \\
    \hat{D}_i&=i\bigg(\ket{u^{J=0}_i}\bra{u^{J=1}}-\ket{u^{J=1}}\bra{u^{J=0}_i}\bigg)
\end{align}

\noindent
We may write these operators in terms of the Pauli matrices as

\begin{align}
    \hat{S}_1&=\dfrac{1}{2\sqrt{2}}(\sigma_0-\sigma_z)\otimes \sigma_x-\dfrac{1}{2\sqrt{2}}\sigma_x\otimes (\sigma_0-\sigma_z) \nonumber \\
    \hat{S}_2&=\dfrac{1}{2}\bigg(\sigma_0\otimes\sigma_z-\sigma_z\otimes\sigma_0\bigg) \nonumber \\
     \hat{S}_3&=-\dfrac{1}{2\sqrt{2}}(\sigma_0+\sigma_z)\otimes \sigma_x+\dfrac{1}{2\sqrt{2}}\sigma_x\otimes (\sigma_0+\sigma_z) \nonumber \\
     \hat{D}_1&=\dfrac{1}{2\sqrt{2}}(\sigma_0-\sigma_z)\otimes \sigma_y-\dfrac{1}{2\sqrt{2}}\sigma_y\otimes (\sigma_0-\sigma_z) \nonumber \\
     \hat{D}_2&=\dfrac{1}{2}\bigg(\sigma_x\otimes\sigma_y-\sigma_y\otimes\sigma_x\bigg) \nonumber \\ 
     \hat{D}_3&=\dfrac{1}{2\sqrt{2}}(\sigma_0+\sigma_z)\otimes \sigma_y+\dfrac{1}{2\sqrt{2}}\sigma_y\otimes (\sigma_0+\sigma_z) 
     \label{matM}
\end{align}

\noindent
Any kinetic hopping proportional to any linear combination of the operators $\hat{S}_i$ and $\hat{D}_i$ will lead to orbital frustration in both multiplets as they are operators that induce intermultiplet hopping for the onsite potential described in equation \ref{intra4dof} and will lead to spin-orbit assisted orbital frustration in both the $J=0$ and $J=1$ multiplets.

\subsection{One $l=1$ and $s=1/2$ Degree of Freedom}
\label{dof6}

Consider a system with unit cells containing six degrees of freedom: an effective $s=1/2$ degree of freedom and an effective $l=1$ orbital degree of freedom.  As described in section \ref{multiplet} onsite spin-orbit interaction will split these degrees of freedom into a twofold degenerate effective $J=1/2$ multiplet and a fourfold degenerate effective $J=3/2$ multiplet whose eigenstate structure takes a similiar form to that in equations \ref{eigstates1} and \ref{eigstates2}.

We can determine the allowed intercell kinetic hopping coefficients that preserve orbital frustration in the presence of a spin-orbit potential of the form $\Lambda^{\alpha\beta}=\lambda_{so} \bra{\alpha} \hat{\bm{L}}\cdot \hat{\bm{S}} \ket{\beta}$ by first decomposing $\hat{H}'_{\text{inter}}(\bm{k})$ as

\begin{equation}
    \bra{\alpha}\hat{H}'_{\text{inter}}(\bm{k})\ket{\beta}=\sum_{i=0}^8\sum_{J=0}^3 t_{ij}(\bm{k}) (\lambda_i\otimes \sigma_j)_{\alpha\beta}
\end{equation}

\noindent
where $\lambda_i$ and $\sigma_j$ are the Gell-Mann and Pauli matrices with $\sigma_0=\mathds{1}_{2\times 2}$ and $\lambda_0=\mathds{1}_{3\times 3}$.  We can then decompose $\hat{H}'_{\text{inter}}(\bm{k})$ into its inter-multiplet and intra-multiplet contributions to determine which terms $t_{ij}(\bm{k})$ could lead to orbital frustration.  In the $J=3/2$ sector all terms $t_{ij}(\bm{k})$ lead to $\mathcal{W}^{J=3/2}_{ij}\neq 0$ and orbital frustration can only occur if a fined tuned linear combination of $t_{ij}(\bm{k})$ is engineered.  However, in the $J=1/2$ sector there are many allowed terms that would still lead to the presence of orbital frustration

\begin{align}
    \text{Allowed}=&\{
    \lambda_1\otimes \sigma_0,
    \lambda_1\otimes \sigma_z,
    \lambda_2\otimes \sigma_x,\nonumber \\
    &\lambda_2\otimes \sigma_y,
    \lambda_3\otimes \sigma_0,
    \lambda_3\otimes \sigma_z,  \nonumber \\
    &\lambda_4\otimes \sigma_0,
    \lambda_4\otimes \sigma_y,
    \lambda_5\otimes \sigma_x, \nonumber \\
    &\lambda_5\otimes \sigma_z,
    \lambda_6\otimes \sigma_0,
    \lambda_6\otimes \sigma_x, \nonumber \\
    &\lambda_7\otimes \sigma_y,
    \lambda_7\otimes \sigma_z,
    \lambda_8\otimes \sigma_0\}
    \label{allow6}
\end{align}

\noindent
In the presence of these terms $\mathcal{W}_{ij}^{J=1/2}=0$ regardless of the functions $t_{ij}(\bm{k})$.  Most of these terms describe an interorbital process by which the $l=1$
orbital degree of freedom is changed upon hopping between sites.  The exceptions are the terms $\lambda_3\otimes \sigma_0$, $\lambda_3\otimes \sigma_z$, and $\lambda_8\otimes \sigma_0$ that all describe a particular combination of intra-orbital hopping by which the orbital and spin degrees of freedom are unchanged upon hopping between sites, but pick up a phase such that the processes destructively interfere and lead to $\mathcal{W}^{J=1/2}_{ij}= 0$.

\section{Lattice Considerations}

Up to this point we have determined the absence or existence of orbital frustration from a local perspective by analyzing various different intracell potentials, their multiplets, and the intercell hopping between these multiplets without any consideration of the crystal momentum dependent functions in $\hat{H}'_{\text{inter}}(\bm{k})$ that contain information about the allowed lattice harmonic functions that can exist on a particular lattice.  Here we consider some simple lattice examples and calculate the band dispersion to analytically show the absence or existence of orbital frustration.

\subsection{Bipartite Lattices}

The examples in section \ref{models}  have demonstrated the importance of particular inter-orbital kinetic structures that give rise to orbital frustration.  Usually inter-orbital coupling can be dominant in multi-partite lattices where nearest neighbor orbitals on the lattice are in-equivalent.  When the Hamiltonian is expressed in terms of a local orbital structure the hopping functions $t^{\alpha\beta}(\bm{R}_i-\bm{R}_j)$ usually are exponentially localized 

\begin{equation}
    t^{\alpha\beta}(\bm{R}_i-\bm{R}_j)\sim e^{-\delta|\bm{R}_i-\bm{R}_j|^\gamma}
\end{equation}

\noindent
for some $\delta,\gamma>0$.  In these cases the largest contribution to $\hat{H}'_{\text{inter}}(\bm{k})$ is  determined by any lattice site's nearest neighbor geometry.

\begin{figure}[!htb]
    \centering
    \includegraphics[width=0.49\textwidth]{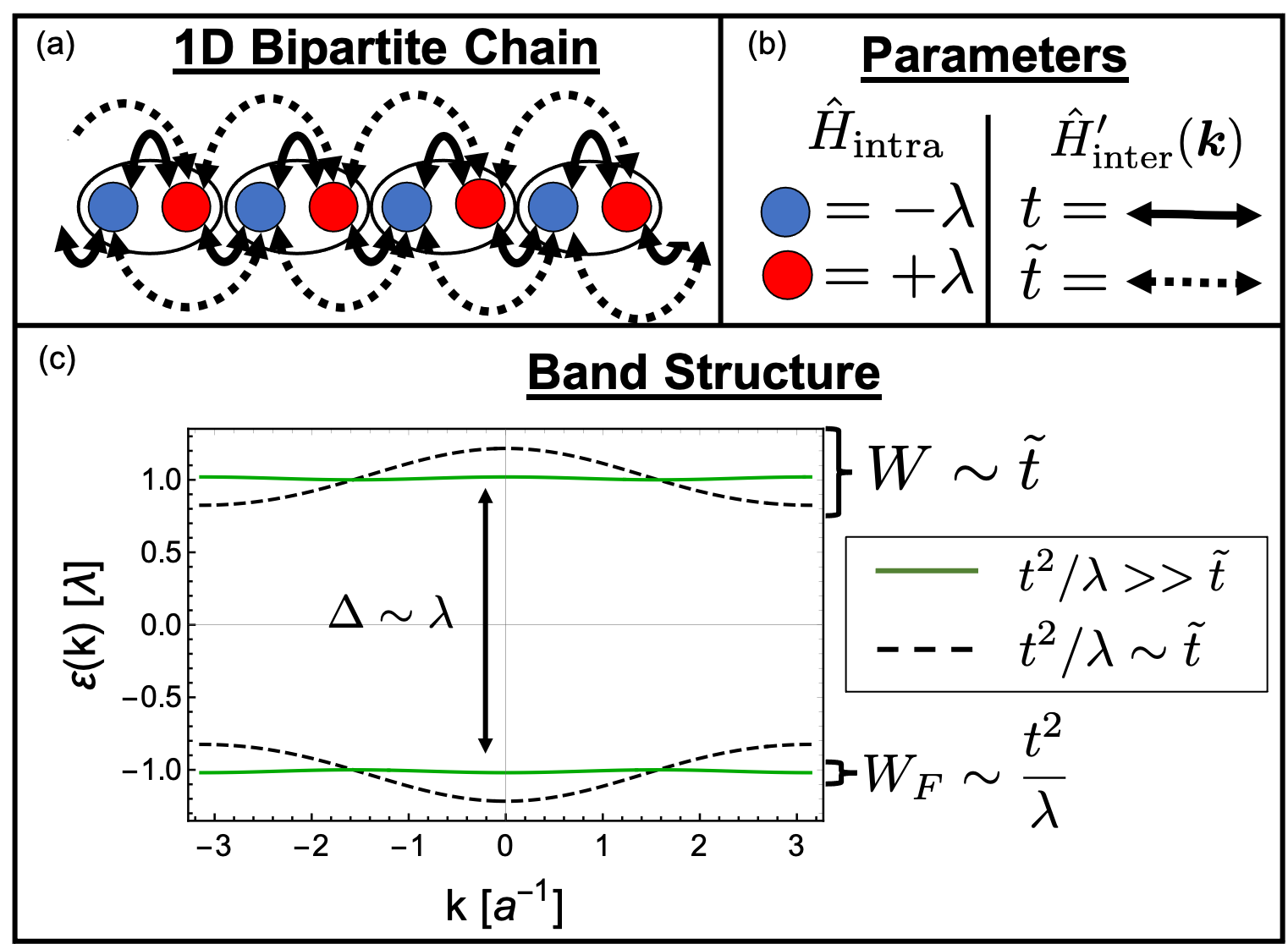}
    \caption{(a) Tight binding model on a 1D bipartite chain. (b) Unit cell contains two orbitals (red and blue) with onsite potentials $\pm\lambda$. Nearest neighbor $t$ and next nearest neighbor $\tilde{t}$ interactions are depicted by double headed arrows. (c) Band structure in the presence ($t=0.1\lambda$, $\tilde{t}=0$, green) and absence ($t=0.1\lambda$, $\tilde{t}=10 t^2/\lambda$, dashed line) of orbital frustration. In the presence of orbital frustration $t^2/\lambda>>\tilde{t}$ and both bands have bandwidths $W_F\sim t^2/\lambda$, while in the absence of orbital frustration $t^2/\lambda\sim\tilde{t}$ and both bands have bandwidths $W\sim \tilde{t}$.}
    \label{fig1D}
\end{figure}

Take as an example the bipartite 1D chain shown in Fig. \ref{fig1D}.  The nature of the different molecular or orbital structures on the $A$ and $B$ sublattice can naturally induce an onsite potential difference that can be described in the Bloch Hamiltonian by an intracellular term $\hat{H}_{\text{intra}}' \sim \hat{\sigma}_z$.  The kinetic dynamics will be dominated by the nearest neighbor interactions that describe a hopping process that takes quanta localized on the $A$ sublattice to the $B$ sublattice and vice versa.  The full Bloch hamiltonian with real hopping processes can be written as

\begin{align}
\hat{H}(k)&=\left(\begin{array}{cc}
   \lambda  & t(1+e^{-ika}) \\
   t(1+e^{ika})  & -\lambda
\end{array}\right) \nonumber \\
&=\bigg(\bm{t}(k)+\bm{\lambda}\bigg)\cdot \hat{\bm{\sigma}}
\end{align}

\noindent
where $a$ is the lattice constant and $\bm{t}(k)=(t+t\cos(ka),t\sin(ka),0)$ and $\bm{\lambda}=(0,0,1)$.  The band eigenvalues are

\begin{equation}
    \varepsilon_\pm(k)=\pm\sqrt{\lambda^2+4t^2\cos^2(ka/2)}
    \label{banddis1}
\end{equation}

\noindent
Due to the orthogonality of the vectors $\bm{t}(k)$ and $\bm{\lambda}$ we see that $\mathcal{W}^\pm(k)=0$ (see equation \ref{w2dof}) and both bands should exhibit orbital frustration.  This can be seen by expanding equation \ref{banddis1} in powers of $t/\lambda$

\begin{equation}
    \varepsilon_\pm(k)\approx \pm|\lambda|\pm \dfrac{2t^2}{|\lambda|}\cos(ka/2)+\mathcal{O}\bigg(\dfrac{t^2}{\lambda^2}\bigg)
\end{equation}

\noindent
and noting that the order $t$ contribution to $\varepsilon_\pm(k)$ is zero.  The bandwidths are of order $W_F\sim t^2/\lambda$ which in the limit of $\lambda\ll t$ is much smaller then the expected bandwidth of order $t$ that would occur in the absence of orbital frustration.  In the presence of next nearest neighbor hopping, $\tilde{t}$, the Hamiltonian will contain processes that take quanta from an $A$ $(B)$ site in one unit cell to a $A$ $(B)$ in another unit cell.  In general these processes will be proportional to $\mathds{1}$ and $\hat{\sigma}_z$ and will spoil orbital frustration if the magnitudes of $\tilde{t}$ is comparable to $t^2/\lambda$.  With the inclusion of next nearest neighbor couplings proportional to $\hat{\sigma}_z$ the Hamiltonian can be written as

\begin{equation}
\hat{H}(k)=\left(\begin{array}{cc}
   \lambda + 2\tilde{t}\cos(ka) & t(1+e^{-ika}) \\
   t(1+e^{ika})  & -\lambda - 2\tilde{t}\cos(ka)
\end{array}\right)
\end{equation}

\noindent
and the band eigenvalues are 

\begin{equation}
     \varepsilon_\pm(k)=\pm\sqrt{(\lambda+2\tilde{t}\cos(ka))^2+4t^2\cos^2(ka/2)}
\end{equation}

\noindent
Here  $\mathcal{W}^\pm(k)=2\tilde{t}\cos(ka)$ and is non-vanishing.  For $t\sim\tilde{t}$ orbital frustration is destroyed and the bandwidths of $\varepsilon_{\pm}(k)$ are of order $W\sim\tilde{t}\sim t$.  As described above in most systems the hopping integrals $t^{\alpha\beta}(\bm{R}_i-\bm{R}_j)$ are exponentially decaying functions of $\bm{R}_i-\bm{R}_j$ such that nearest neighbor interactions are usually the most dominant kinetic coupling in $\hat{H}$.  We see that in the limit of $\tilde{t}\ll t^2/\lambda$ orbital frustration persists ($W_F\sim t^2/\lambda$) as the dominant contribution to the bandwidth derives from the frustrated processes that lead to the vanishing of $\mathcal{W}^\pm(\bm{k})$ in the absence of $\tilde{t}$.  Fig. \ref{fig1D}c shows the band structure of the 1D bipartite chain in the presence ($\tilde{t}\ll t^2/\lambda$) and absence ($\tilde{t}\gtrsim t^2/\lambda$) of orbital frustration.

Similarly, in two dimensions, on the honeycomb lattice, a strong sublattice symmetry breaking potential $\lambda$ and nearest neighbor interaction $t$ will lead to orbital frustration.  The full Bloch Hamiltonian is

\begin{equation}
    \hat{H}(\bm{k})=\left(\begin{array}{cc}
       \lambda  & tf_H(\bm{k}) \\
        tf_H^*(\bm{k}) & -\lambda
    \end{array}\right)
\end{equation}

\noindent
with $f_H(\bm{k})=(1+e^{-i\bm{k}\cdot\bm{R}_1}+e^{-i\bm{k}\cdot\bm{R}_2})$ where $\bm{R}_1$ and $\bm{R}_2$ are the primitive honeycomb lattice vectors $\bm{R}_1=a(1/2,\sqrt{3}/2)$ and $\bm{R}_1=a(-1/2,\sqrt{3}/2)$.  The eigenvalues to leading order in $t$ are

\begin{gather}
    \varepsilon_\pm(\bm{k})\approx\pm|\lambda|\pm \dfrac{t^2}{2|\lambda|}|f_H(\bm{k})|^2
\end{gather}

\noindent
The order $t$ contribution for both band eigenvalues $\varepsilon_\pm(\bm{k})$ vanish, and as a result, $\mathcal{W}^\pm(\bm{k})=0$ and both bands are orbitally frustrated leading to bandwidths of order $W_F\sim t^2/\lambda$.  Again the inclusion of longer range hopping processes will not destroy the frustration as long as the dominant kinetic hopping is still from the nearest neighbor interaction $t$.  This model has been studied in the $\lambda\gg t$ limit in the presence of strong electron-electron interactions where it has been shown that pairing between electrons can be induced by multi-particle tunneling processes between the two polarized electronic states each localized on one of the sublattice sites of the honeycomb and as a result the superconducting transition temperature $T_c$ shows strong dependence on the ratio $t/\lambda$ in this strongly orbitally frustrated system \cite{crepel2020exact}.

\subsection{The Lieb Lattice}

Here we present a special example of spin-orbit assisted orbital frustration whereby the kinetic hopping elements are completely frustrated in a sector of the Hilbert space resulting in a band structure consisting of a perfectly flat band with $\varepsilon(\bm{k})=0$.

The Lieb lattice model describes a tight binding Hamiltonian of nearest neighbor interactions on the square lattice with three atoms per unit cell.  The Bloch Hamiltonian takes the form

\begin{equation}
    \hat{H}(\bm{k})=\left(\begin{array}{ccc}
      0  & \lambda + te^{-i a k_x} & \lambda + te^{-i a k_y} \\
      \lambda + te^{i a k_x}  & 0 & 0 \\
      \lambda + te^{i a k_y}  & 0 & 0 
    \end{array}
    \right)
\end{equation}

\noindent
whose eigenvalues are

\begin{align}
    \varepsilon_0(\bm{k})&=0 \nonumber \\
    \varepsilon_0(\bm{k})&=\pm \sqrt{2}\sqrt{t^2+\lambda^2+t\lambda(\cos(k_x)+\cos(k_y))}
\end{align}

\noindent
Here $\lambda$ denotes the hopping within the unit cell and $t$ the hopping between unit cells.  The flat band $\varepsilon_0(\bm{k})$ can be understood in the terms of spin-orbit assisted orbital frustration as follows.

The Hamiltonian can be written as

\begin{align}
\hat{H}(\bm{k})&= \hat{H}'_{\text{intra}}+\hat{H}'_{\text{inter}}(\bm{k}) \nonumber  \\
\hat{H}'_{\text{intra}}&=\left(\begin{array}{ccc}
      0  & \lambda  & \lambda \\
      \lambda  & 0 & 0 \\
      \lambda  & 0 & 0 
    \end{array}
    \right) \nonumber \\
\hat{H}'_{\text{inter}}(\bm{k})&= \left(\begin{array}{ccc}
      0  & e^{-i a k_x}  & e^{-i a k_y} \\
      e^{i a k_x}  & 0 & 0 \\
      e^{i a k_y}  & 0 & 0 
    \end{array}
    \right)
\end{align}

\noindent
The eigenstate of $\hat{H}'_{\text{intra}}$ are

\begin{align}
    \ket{u_0}&=\dfrac{1}{\sqrt{2}}(0,-1,1) \nonumber \\
    \ket{u_\pm}&=\dfrac{1}{\sqrt{2}}(\pm\sqrt{2},1,1) \nonumber \\
\end{align}

\noindent
Here the multiplet of interest is singly degenerate such that the first order correction to the eigenvalues is simply given by 

\begin{equation}
\mathcal{W}^0(\bm{k})=\bra{u_0}\hat{H}'_{\text{inter}}(\bm{k})\ket{u_0}=0
\end{equation}

\noindent
The vanishing of $\mathcal{W}^0(\bm{k})$ signifies orbital frustration as this Lieb lattice model only allows kinetic processes that connect eigenstates in different multiplets.

It turns out that higher order corrections to the energy eigenvalues in the Lieb lattice also vanish.  At second order in $\lambda$ this can be seen by computing $\varepsilon_0^{(2)}(\bm{k})$ using equation \ref{corr2}.  A general proof of the existence of a flat band for all values of $\lambda,t$ can be determined using the S-matrix techniques developed in references \cite{li2021catalogue,cualuguaru2022general}.

\subsection{Larger Degrees of Freedom Per Unit Cell}

Next we examine orbital frustration in a model of four degrees of freedom per-site on a primitive square lattice.  We take the onsite potential to be the spin-spin like interaction described in equation \ref{intra4dof} with $\lambda_0>0$.  Any terms in the intercell Hamiltonian proportional to the allowed terms of equation \ref{halfhalfOF} lead to orbital frustration in the $J=0$ multiplet of states.  As an example, here we take the nearest neighbor intercell hopping

\begin{equation}
    \bra{\alpha}\hat{H}'_{\text{inter}}(\bm{k})\ket{\beta}=tf_S(\bm{k})(\sigma_y \otimes \sigma_z)_{\alpha\beta}
\end{equation}

\noindent
where $f_S(\bm{k})=2t(\cos(k_x a)+\cos(k_y a))$.  The eigenvalues to lowest order in $t$ are

\begin{align}
    \varepsilon_{J=0}(\bm{k})&\approx-3\lambda_0-\dfrac{t^2}{4\lambda_0}(f_S(\bm{k}))^2 \nonumber \\
    \varepsilon^1_{J=1}(\bm{k})&\approx \lambda_0-tf_S(\bm{k}) \nonumber \\
    \varepsilon^2_{J=1}(\bm{k})&\approx\lambda_0+\dfrac{t^2}{4\lambda_0}(f_S(\bm{k}))^2 \nonumber \\
    \varepsilon^3_{J=1}(\bm{k})&\approx \lambda_0+tf_S(\bm{k}) 
\end{align}

\begin{figure*}[!htb]
    \centering
    \includegraphics[width=0.99\textwidth]{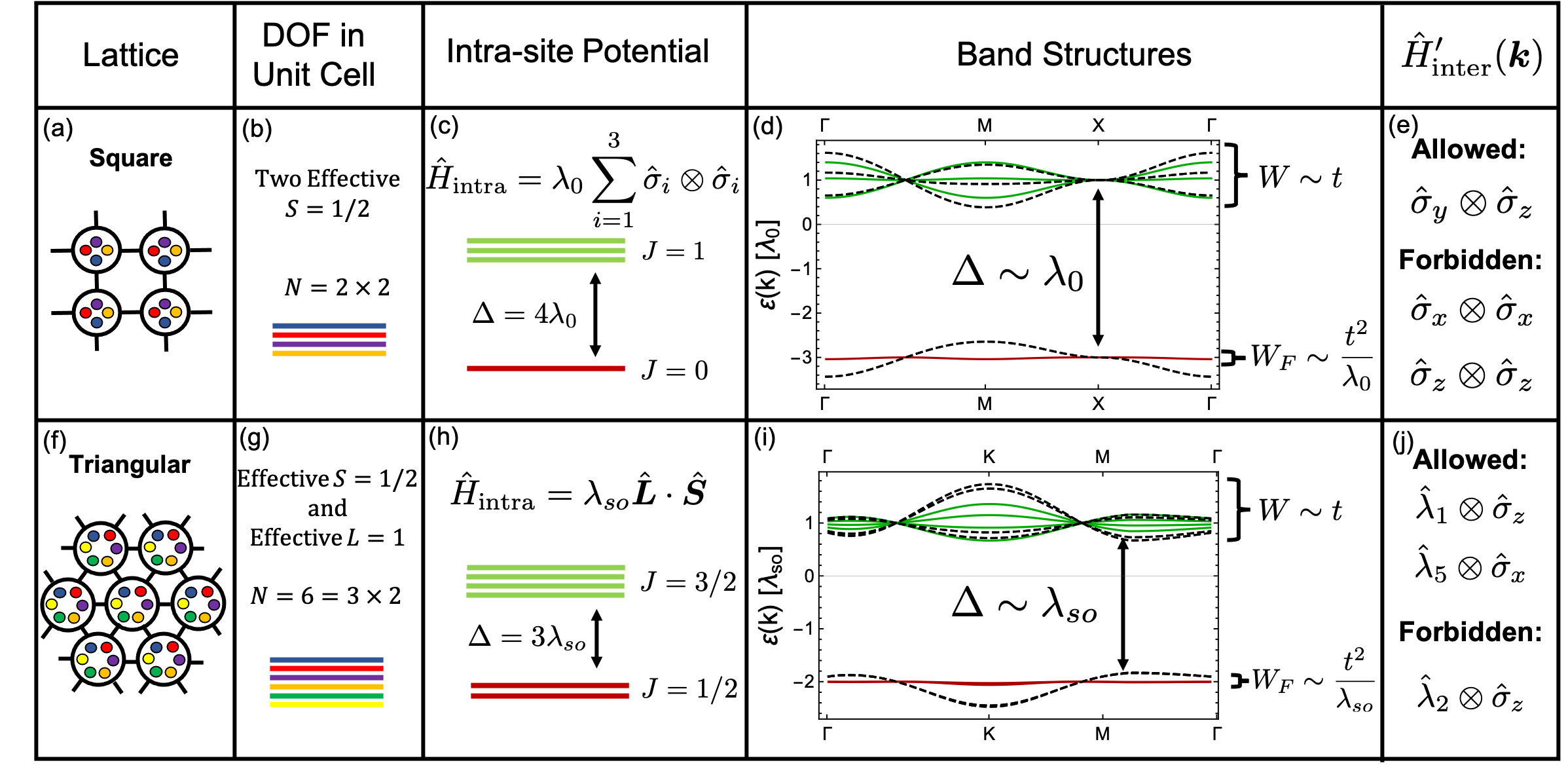}
    \caption{Examples of orbital frustration in two multi-orbital two dimensional models. (a)-(e) Square lattice model (a) with two effective spin 1/2 degrees of freedom per unit cell (b) with intra-site potential splitting the orbitals into $J=1$ and $J=0$ multiplets (c).  (d) Band structure in the presence and absence of orbital frustration for intercell kinetic couplings shown in (e) with $t=0.1\lambda_0$.  (d) In the presence of allowed terms (solid lines) the $J=0$ multiplet is frustrated and has a bandwidth $W_F\sim t^2/\lambda_0$, while in the presence of the forbidden terms in (e) orbital frustration is spoiled in both multiplets leading to bandwidths $W\sim t$.  (f)-(j) Triangular lattice model with effective $L=1$ orbital and effective $S=1/2$ spin degrees of freedom per unit cell (g) coupled by an intra-site potential splitting the degrees of freedom into a $J=3/2$ and $J=1/2$ multiplet structure (h).  (i) Band structure in the presence and absence of orbital frustration with $t=0.1\lambda_{so}$.  In the presence of allowed terms in (j) the $J=1/2$ multiplet is frustrated and has bandwidth $W_F\sim t^2/
    \lambda_{so}$ (solid lines in (i)), while in the presence of forbbiden terms in (j) orbital frustration is spoiled in both multiplets (dashed-lines in (i)) leading to bandwidths $W\sim t$. }
    \label{figSO}
\end{figure*}

\noindent
We see orbital frustration in the $J=0$ multiplet marked by the vanishing of the order $t$ correction to the band energies of $\hat{H}'_{\text{intra}}$ and the vanishing of equation \ref{w4dof}.  While in the $J=1$ multiplet 

\begin{gather}
    \mathcal{W}^{J=1}(\bm{k})=f_S(\bm{k})
    \left(\begin{array}{ccc}
        0 & -i/\sqrt{2} & 0  \\
        i/\sqrt{2} & 0 & i/\sqrt{2} \\
        0 & -i/\sqrt{2} & 0
    \end{array}\right)
\end{gather}

\noindent
whose eigenvalues are $\pm 1$ and $0$.  The zero eigenvalue marks the existence of orbital frustration in $\varepsilon^2_{J=1}(\bm{k})$, while the eigenvalues $\pm 1$ determines the absence of orbital frustration in $\varepsilon^1_{J=1}(\bm{k})$ and $\varepsilon^3_{J=1}(\bm{k})$ (see Fig. \ref{figSO}(a)-(e)).  

As shown in section \ref{4DOF} to achieve orbital frustration in both the $J=1$ and $J=0$ multiplets the intercell kinetic hopping needs to be proportional to some linear combination of the matrices in equation \ref{matM}.  As an example consider

\begin{align}
    \hat{H}'_{\text{inter}}(\bm{k})&=tf_S(\bm{k})(\hat{S}_1+\hat{S}_3) \nonumber \\
    &=\dfrac{t}{\sqrt{2}}f_S(\bm{k})(\hat{\sigma}_x\otimes \hat{\sigma}_z-\hat{\sigma}_z\otimes\hat{\sigma}_x)
\end{align}

\noindent
The eigenvalues to leading order in $t$ are

\begin{align}
    \varepsilon_{J=0}&\approx-3\lambda_0 -\dfrac{t^2}{2\lambda}(f_S(\bm{k}))^2\nonumber \\
    \varepsilon^1_{J=1}&=\lambda_0 \nonumber \\
    \varepsilon^2_{J=2}&=\lambda_0 \nonumber \\
    \varepsilon^3_{J=3}&\approx\lambda_0 +\dfrac{t^2}{2\lambda}(f_S(\bm{k}))^2
\end{align}

\noindent
In this special case the matrix $\tilde{\mathcal{W}}_{ij}^{nm}=\bra{u^n_i}\hat{H}'_{\text{inter}}\ket{u^m_j}$ has two zero eigenvalues whose eigenvectors are linear combinations of states in the $J=1$ multiplet. This is true for any linear combination of $\hat{S}_i$ and $\hat{D}_i$ and will lead to complete orbital frustration where two bands in the multiplet are completely unperturbed by $\hat{H}'_{\text{inter}}(\bm{k})$ and as such remain completely flat across the Brillouin zone in the presence of $\hat{H}'_{\text{inter}}(\bm{k})$.

Lastly we consider a triangular lattice with an effective spin $1/2$ and effective spin $1$ degree of freedom coupled by the onsite spin-orbit potential describe in equation \ref{LScouple} (see Fig. \ref{figSO}).  For simplicity we take an intercell potential of the form

\begin{equation}
    \bra{\alpha}\hat{H}'_{\text{inter}}(\bm{k})\ket{\beta}=f_T(\bm{k})\bigg(t_{13}\hat{\lambda}_1\otimes \hat{\sigma}_z+t_{51}\hat{\lambda}_5\otimes \hat{\sigma}_x\bigg)_{\alpha\beta}
\end{equation}

\noindent
with
\begin{gather}
    f_T(\bm{k})=2\cos(k_xa)+4\cos\bigg(\dfrac{\sqrt{3}}{2}k_ya\bigg)\cos\bigg(\dfrac{k_x a}{2}\bigg)
\end{gather}

\noindent
where we have chosen two particular kinetic coupling from equation \ref{allow6}. As such, $\mathcal{W}^{J=1/2}=0$ and $\mathcal{W}^{J=3/2}\neq0$ and there will be orbital frustration in the $J=1/2$ multiplet of bands, but not the $J=3/2$ multiplet.  This is reflected in the eigenvalues of the full Bloch Hamiltonian.  For example, if we take $t_{13}=t_{51}/2=t/3$ and $\lambda_{so}>0$ the eigenvalues to leading order in $t$ are

\begin{align}
    \varepsilon^1_{J=1/2}(\bm{k})&\approx-2\lambda_{so}-\dfrac{2t^2}{27\lambda}(f_T(\bm{k}))^2 \nonumber \\
    \varepsilon^2_{J=1/2}(\bm{k})&\approx-2\lambda_{so}-\dfrac{14t^2}{81\lambda}(f_T(\bm{k}))^2 \nonumber \\
    \varepsilon^1_{J=3/2}(\bm{k})&\approx\lambda_{so}-\dfrac{t}{\sqrt{3}}f_T(\bm{k})) \nonumber \\
    \varepsilon^2_{J=3/2}(\bm{k})&\approx\lambda_{so}-\dfrac{t}{3\sqrt{3}}f_T(\bm{k}) \nonumber \\
    \varepsilon^3_{J=3/2}(\bm{k})&\approx\lambda_{so}+\dfrac{t}{3\sqrt{3}}f_T(\bm{k}) \nonumber \\
    \varepsilon^3_{J=3/2}(\bm{k})&\approx\lambda_{so}+\dfrac{t}{\sqrt{3}} f_T(\bm{k})\nonumber
\end{align}

\noindent
The leading order contribution to the eigenvalues of the $J=1/2$ multiplet are of order $t^2/\lambda$ marking the presence of orbital frustration, while in the $J=3/2$ multiplet the leading order in $t$ is linear for all eigenvalues.  Fig. \ref{figSO}(i) shows the band structure in the limit of $t\gg\lambda$ along the high symmetry lines of the Brillouin zone in the pressence and absence of spin-orbit assisted orbital frustration.

\section{Symmetry Considerations}

The symmetry of the system further constrains the allowed inter-cellular hopping terms that can appear in $\hat{H}'_{\text{inter}}(\bm{k})$.  Here we present a model of nearest-neighbor interacting spin-1/2 $p$-orbitals on a square lattice whose degrees of freedom in a unit cell are coupled by a large intra-cellular spin-orbit interaction $\hat{H}'_{\text{intra}}=\lambda_{so} \bm{L}\cdot\bm{S}$.  As described in section \ref{multiplet} the strong intra-cellular potential will split the degrees of freedom into a two fold degenerate effective $J=1/2$ multiplet and a four fold degenerate effective $J=3/2$ multiplet.  

In section \ref{dof6} we showed the allowed matrix structure in $\hat{H}'_{\text{inter}}(\bm{k})$ that would lead to an orbital frustrated $J=1/2$ multiplet, however, the square lattice contains additional symmetries that if preserved put restrictions on $\hat{H}'_{\text{inter}}(\bm{k})$.  Each site has four nearest neighbors: two that form bonds in the $\hat{\bm{x}}$-direction and two that form bonds in the $\hat{\bm{y}}$-direction. The symmetries $\mathcal{O}$ of interest leave the bond vector from site $\bm{R}_i$ to $\bm{R}_j$, $\bm{R}_i-\bm{R}_j$, invariant ($\mathcal{O}(\bm{R}_i-\bm{R}_j)=(\bm{R}_i-\bm{R}_j)$).  This puts constraints on the values $t_{ij}^{\alpha\beta}$ via

\begin{equation}
   \sum_\gamma \bigg( \mathcal{O}_{\alpha\gamma}t_{\gamma\beta}^{ij}-t_{\alpha\gamma}^{ij}\mathcal{O}_{\gamma\beta}\bigg)=0
   \label{symcont}
\end{equation}

\noindent
For the $x$-bonds these symmetries are mirror-z, $\mathcal{M}_z$, about the material plane containing the bond vectors and mirror-y, $\mathcal{M}_y$, about the plane intersecting the bond.  In the local orbital basis spanned by the indices $\alpha,\beta$ the symmetry operators in equation \ref{symcont} can be written as

\begin{align}
   (\mathcal{M}_y)_{\alpha\beta}&=\bigg((\lambda_3+(\mathds{1}-\sqrt{3}\lambda_8)/3) \otimes-i\sigma_y\bigg)_{\alpha\beta}\nonumber \\
    (\mathcal{M}_z)_{\alpha\beta}&=\bigg((\mathds{1}+\sqrt{3}\lambda_8)\otimes-i\sigma_z\bigg)_{\alpha\beta}
\end{align}

\noindent
We note that two-fold rotations along the bond axes $C_{2x}$ satisfy $C_{2x}=\mathcal{M}_y\mathcal{M}_z$.  The allowed hopping in the $\hat{\bm{y}}$-direction can be determined by a fourfold rotation about the $z$-axis.  We further constrain the Hamiltonian by assuming time reversal symmetry $\hat{\mathcal{T}}=i\sigma_y \hat{\mathcal{K}}$, where $\mathcal{K}$ is the complex conjugation operator.

This leads to six independent hopping parameters $(t_{S1},t_{S2},t_{S3},t_{S4},t_{D1},t_{D2})$ whose Bloch Hamiltonian in the basis $(\ket{p_x,\uparrow},\ket{p_y,\uparrow},\ket{p_z,\uparrow},\ket{p_x,\downarrow},\ket{p_x,\downarrow},\ket{p_x,\downarrow})$ we write as

\begin{equation}
    \hat{H}'_{\text{inter}}(\bm{k})=\left(\begin{array}{cc}
      H_{S}(\bm{k})   & H_{D}(\bm{k})  \\
        (H_{D}(\bm{k}))^\dagger & H_{S}^*(\bm{k}) 
    \end{array}\right)
\end{equation}

\noindent
where the $S$ and $D$ denote couplings between same and different spin characters.  Here

\begin{equation}
H_S(\bm{k})=\left(\begin{array}{ccc}
    F(t_{S1},t_{S2}) & iG(t_{S4}) & 0 \\
    -iG(t_{S4}) &  F(t_{S2},t_{S1}) & 0 \\
    0 & 0 & G(t_{S3}) 
\end{array}    \right)
\end{equation}

\noindent
and 

\begin{equation}
H_D(\bm{k})=\left(\begin{array}{ccc}
   0 & 0 & F(t_{D1},t_{D2}) \\
    0 &  0 & iF(t_{D2},t_{D1}) \\
    -F(t_{D1},t_{D2}) & -iF(t_{D2},t_{D1}) & 0
\end{array}    \right)
\end{equation}

\noindent
with

\begin{align}
    F(t_1,t_2)&=2(t_1\cos(k_x)+t_2\cos(ky)) \nonumber \\
    G(t)&=2t(\cos(k_x)+\cos(ky))
\end{align}

To derive the constraint for orbital frustration in the $J=1/2$ multiplet we construct $\mathcal{W}^{J=1/2}_{ij}$ using the eigenfunctions of $\hat{H}'_{\text{intra}}=\lambda_{so} \bm{L}\cdot\bm{S}$ which take the same form as those given in equation \ref{eigstates1}.  We find that $\mathcal{W}^{J=1/2}_{ij}$ vanishes when the following condition is satisfied:

\begin{equation}
    2(t_{D1}+t_{D2}+t_{S4})- (t_{S1}+t_{S2}+t_{S3})=0
    \label{constraint}
\end{equation}

\noindent
If equation \ref{constraint} is satisfied the degrees of freedom of the effective $J=1/2$ multiplet will exhibit orbital frustration and disperse with a narrow bandwidth of order $t^2/\lambda_{so}$.

\begin{figure}[!htb]
    \centering
    \includegraphics[width=0.45\textwidth]{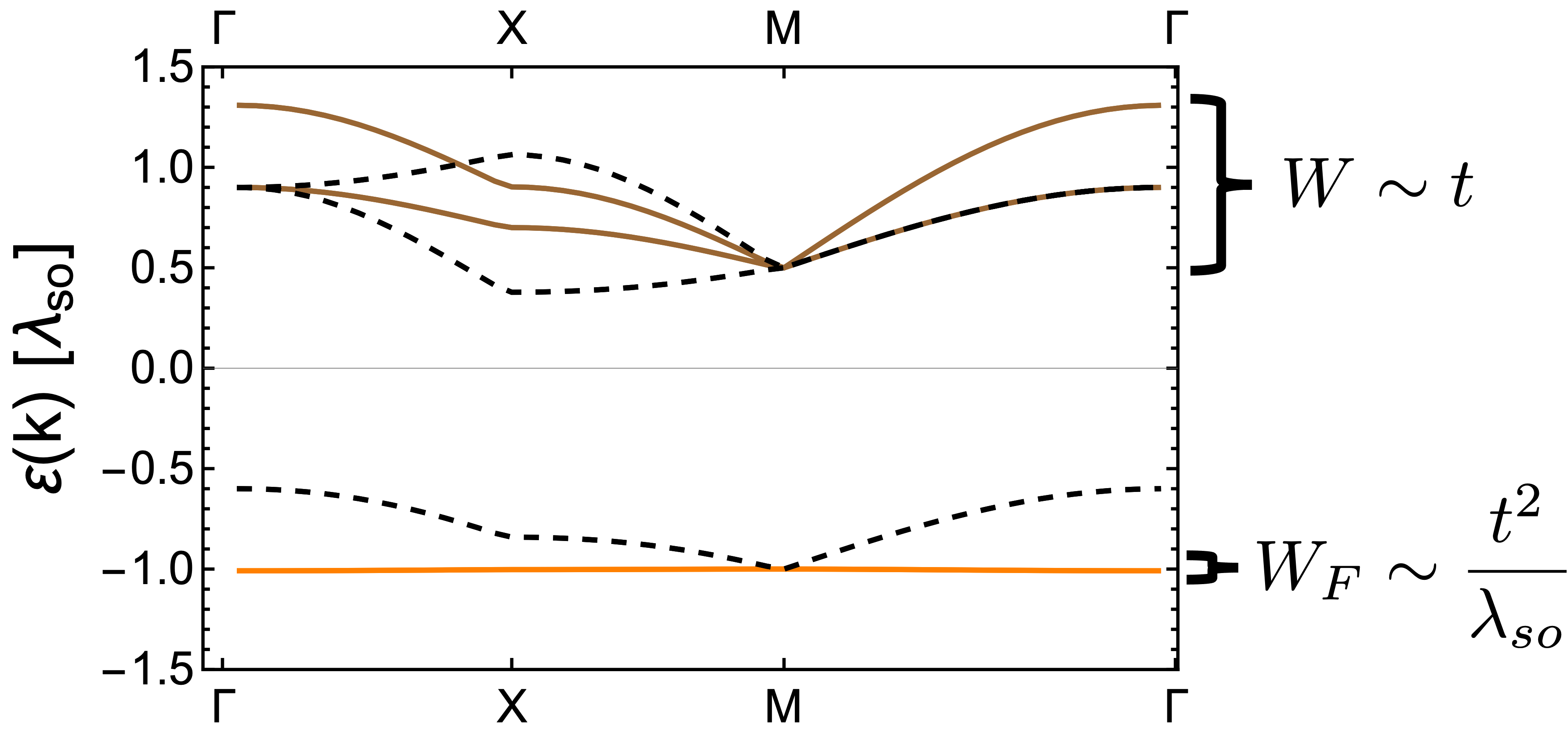}
    \caption{Spin-orbit assisted orbital frustration in a model of nearest-neighbor interacting spin-1/2 $p$-orbitals on a square lattice whose degrees of freedom in a unit cell are coupled by a large intra-cellular spin-orbit interaction $\hat{H}_{\text{intra}}=\lambda_{so} \bm{L}\cdot\bm{S}$.  Band structure in the presence of orbital frustration ($t_{S1}=t_{S2}=t_{S3}=0.1\lambda_{so}$, $t_{S4}=0$, $t_{D1}=t_{D2}=0.075\lambda_{so}$, solid lines) where the degrees of freedom of the effective $J=1/2$ multiplet (orange) are frustrated.  Dashed lines correspond to a system with $t_{S1}=t_{S2}=t_{S3}=0.1\lambda_{so}$, $t_{S4}=0$, and $t_{D1}=-t_{D2}=0.15\lambda_{so}$, where spin-orbit assisted orbital frustration is absent and both the effective $J=1/2$ and $J=3/2$ multiplet disperse with bandwidths $W\sim t$.  Note that due to Kramer's degeneracy endowed by time-reversal and inversion symmetries all bands shown are two fold degenerate.}
    \label{figSymm}
\end{figure}

Fig. \ref{figSymm} shows the band structure for a system in the absence and presence of spin-orbit assisted orbital frustration.  The solid lines correspond to a system where equation \ref{constraint} is satisfied and for which the degrees of freedom in the effective $J=1/2$ multiplet are frustrated leading to narrowly dispersing bands of bandwidth $t^2/\lambda_{so}$.  The dashed lines correspond to a system in the absence of spin-orbit assisted orbital frustration for which the constraint in equation \ref{constraint} is not satisfied and for which both the effective $J=1/2$ and $J=3/2$ multiplets disperse with bandwidth $W\sim t$.

\section{Conclusion}

Spin-orbit assisted orbital frustration is a new route to engineer and search for flat band systems in a variety of physical settings.  Here we have presented some simple flat band models on 1D and 2D lattices with a varying number of degrees of freedom per unit cell and with different types of intracell potentials that mix and hybridize these degrees of freedom.  Recent work has used density functional theory to search and catalogue over 2,000 candidate flat band materials  \cite{li2021catalogue}.  It is to be seen whether or not some of the proposed materials exhibit spin-orbit assisted orbital frustration.  We propose that by engineering flat bands via the inspired design principle laid out above one can exploit the nature of how these flat bands arise to minimize bandwidths and flatness ratios in a highly controlled manner and to circumvent the difficulties of exploring the innumerably large phase-space of stable material compounds.

Highly tunable platforms, like ultra-cold atoms, photonic crystals, and quantum circuits could provide a platform to realize the models presented above.  Due to the narrow band behavior of these system's multiplets, interactions and non-linearities could play a fundamental role in determining the nature of the ground and excited states of these systems.  It has been shown in a model of $d$-orbitals on a honeycomb lattice that spin-orbit assisted orbital frustrated bands give rise to a purely Kitaev spin liquid in the presence of interactions \cite{zhang2021orbital}. These models can serve as a platform for determining the nature of other strongly interacting states of matter that could arise in these highly frustrated flat band systems in the presence of interactions.  In doing so the study of spin-orbit assisted orbital frustration can be used to understand the relationship between orbitally frustrated flat bands, interactions, and exotic strongly correlated states of matter.

\smallskip
\noindent
{\bf Acknowledgements:} 
Z.A. and N.T. acknowledge support from NSF Materials Research Science and Engineering Center Grants No. DMR-1420451 and DMR-2011876.  Z.A. acknowledges support from The Ohio State President's Postdoctoral Scholars Program sponsored by the Office of the President.

\bibliography{mybib}

\appendix

\end{document}